\documentclass[a4paper,11pt]{article}
\pdfoutput=1 % if your are submitting a pdflatex (i.e. if you have
\usepackage{jcappub} % for details on the use of the package, please
\usepackage[T1]{fontenc} % if needed

\usepackage{url}
\usepackage{color}
\usepackage{comment}
\usepackage{mathrsfs}
\usepackage{hyperref}% add hypertext capabilities
\usepackage{braket}
\usepackage[dvipsnames]{xcolor}

\newcommand{\simgt}{\lower.5ex\hbox{$\; \buildrel > \over \sim \;$}}

\renewcommand{\hrulefill}{\leavevmode\leaders\hrule height 1pt\hfill\kern0pt }

\title{The impact of the redshift-dependent selection effect of halos
on the redshift-space power spectrum }

\author[a,b,c]{Kanmi Nose,}
\author[a,c]{Masahiro Takada,}
\author[a,b,c]{Ryo Terasawa}

\affiliation[a]{Kavli Institute for the Physics and Mathematics of the Universe (WPI), The University of Tokyo Institutes for Advanced Study (UTIAS), The University of Tokyo, 5-1-5 Kashi- wanoha, Kashiwa-shi, Chiba, 277-8583, Japan}
\affiliation[b]{Department of Physics, The University of Tokyo, Bunkyo, Tokyo 113-0031, Japan}
\affiliation[c]{Center for Data-Driven Discovery (CD3), Kavli IPMU (WPI), UTIAS, The University of Tokyo, Kashiwa, Chiba 277-8583, Japan}

\emailAdd{kanmi.nose@ipmu.jp}

\abstract{In a wide-area spectroscopic survey of galaxies, it is nearly impossible to obtain a homogeneous sample of galaxies
with respect to galaxy properties such as 
stellar mass and host halo mass
across a range of redshifts.
Despite the
selection effect, 
theoretical templates in most analyses assume single tracers
when
compared with the measured clustering quantities.
We demonstrate analytically that the selection effect inevitably introduces a bias in the redshift-space power spectrum on scales from linear to nonlinear scales.
To quantitatively assess the impact of the selection effect, we construct mock galaxy catalogs from halos in N-body simulations by selecting halos above redshift-dependent mass thresholds such that the resulting redshift distribution 
of the halos, 
$n(z)$, matches that of SDSS-like galaxies.  
We find that the selection effect causes fractional changes of up to only 1\% and 2\%  in the monopole and quadrupole moments of 
the redshift-space power spectrum at $k\lesssim 0.3~h{\rm Mpc}^{-1}$, respectively,  compared to the moments for the single mass-threshold (therefore single tracer) sample, for $n_{\rm g}(z)$ of the SDSS-like galaxy samples. 
We also argue that the selection effect is unlikely to cause a significant bias in the estimation of cosmological parameters using the Fisher matrix method, provided that the redshift-dependent selection effect is modest. 
}

\begin{document}
\maketitle
\flushbottom

\section{Introduction}\label{sec:intro}
Observations of clustering properties of galaxies on cosmological scales from a wide-area spectroscopic galaxy survey provide a powerful 
way
%of the fundamental physics of the Universe, such as the nature of dark matter and dark energy, the expansion history and the physics of inflation 
%\citep{2013PhR...530...87W,2018PhR...733....1D,2020moco.book.....D}.
%Large-scale structure (LSS) contains much information for the fundamental mysteries in cosmology such as the nature of the dark matter, dark energy and the origin of the universe.
%The standard method to extract the information is to study the statistics of the tracer of LSS.
%For example, the three dimensional distribution of galaxy, measured form wide-area spectroscopic surveys, is one of the powerful tools 
for constraining cosmological parameters, 
%of the standard $\Lambda$CDM model, 
exploring the physics of inflation, and 
%finding properties of the primordial perturbations and 
testing gravity theories on cosmological scales \citep{2013PhR...530...87W,2018PhR...733....1D,2020moco.book.....D}.
%beyond general relativity.
%To achieve these big purposes, 
There are various existing, ongoing and planned galaxy redshift surveys: the SDSS-III Baryon Oscillation Spectroscopic Survey (BOSS) \cite{2013AJ....145...10D}, the SDSS-IV extended Baryon Oscillation Spectroscopic Survey (eBOSS) \cite{2016AJ....151...44D}, the Subaru Prime Focus Spectrograph (PFS) \cite{2014PASJ...66R...1T}, 
the Dark Energy Spectroscopic Instrument (DESI) \cite{2016arXiv161100036D}, the ESA Euclid\footnote{\url{https://www.esa.int/Science_Exploration/Space_Science/Euclid}}, 
%Rubin Observatory's Legacy Survey of Space and Time (LSST) \cite{2011arXiv1110.3193L}, 
and the NASA Nancy Grace Roman 
Space Telescope \cite{2015JPhCS.610a2007G}.

A standard approach to galaxy clustering analysis involves the use of the two-point correlation function or its Fourier-transformed counterpart, 
the power spectrum. 
The baryonic acoustic oscillations (BAO) \citep{Eisenstein:1997ik} and the apparent anisotropic clustering due to the use of an incorrect cosmological model, i.e. the Alcock-Paczynski (AP) effect \citep{Alcock:1979mp}, in the measured two-point statistics provide a powerful geometrical probe of cosmological distances \citep{SDSS:2005xqv,2dFGRS:2005yhx}. 
In addition, peculiar velocities of galaxies cause characteristic anisotropies in redshift-space distribution of galaxies, i.e. redshift-space distortion (RSD) 
effect \citep{1987MNRAS.227....1K}, and can be used to probe the growth of 
structure, free of galaxy bias uncertainty \citep{Peacock_2001,Reid_2012}. 
More recently, advancements in the theory and methods of galaxy clustering analysis have enabled a more comprehensive approach, known as the “full-shape” analysis 
\citep{Ivanov_2020,kobayashi2022fullshapecosmologyanalysissdssiii}. 
This method incorporates BAO, AP, and RSD information as well as the shape information
of the underlying matter power spectrum up to the quasi nonlinear scales
within an assumed theoretical framework, such as the $\Lambda$CDM model.

%The galaxy clustering analysis above often measures the power spectrum, one of the statistics of the galaxy distribution, for the  comparison with theoretical templates.
%This is because this statistics has characteristic features useful for tight cosmological constraints.
%In particular, the line-of-sight peculiar velocities of the galaxies cause Doppler effect and modulate the intrinsic redshifts of the individual galaxies.
%This induces an apparent anisotropy in the galaxy power spectrum even the cosmological principle states the homogeneous and isotropic universe, and this effect is so-called redshift-space distortion (RSD) \citep{1984ApJ...284L...9K,2000MNRAS.312..257H}.
%Since RSD is a gravitational effect, it enable us to test gravity theory on cosmological scale \cite{2000MNRAS.313..141V} as well as constrain the cosmological parameters.

However, the unknown relationship between the galaxy and 
%underlying 
dark matter distributions in large-scale structure, known as galaxy bias uncertainty, is 
%has been 
a main obstacle to galaxy clustering
cosmology.
% based on galaxy clustering.
Galaxy bias varies among different types of galaxies, depending on factors such as stellar mass, host halo mass, and star formation history, in a complex manner \citep[e.g.][]{Springel_2017}.
Given this fact, it is very difficult or nearly impossible to construct a {\em homogeneous} sample of galaxies from galaxy survey data according to the same selection cuts of observables such as their stellar masses across a range of redshifts.
For example, for galaxies at higher redshifts, it is easier to observe intrinsically brighter galaxies for the same exposure time. 
In addition, cuts on colors and apparent magnitudes are often applied to select target galaxies for spectroscopic observations \citep[e.g.][]{2001AJ....122.2267E,2016MNRAS.455.1553R}, where
 galaxy colors and magnitudes are redshift-dependent.
 These result in a non-trivial and inhomogeneous selection of spectroscopic galaxies across redshifts, which we hereafter refer to as the ``selection effect''.
 Although the selection effect is unavoidable in the data, most cosmological analyses  adopt theoretical templates of single tracers, effectively ignoring the selection effect, to estimate cosmological parameters by comparing with the measurements.

%In general, this relation depends on types of galaxies such as color and magnitude, but can not be modeled completely from the first principle because the physics in galaxy formation and evolution are too complicated.
%Therefore this biased relation prevents us from constructing a perfect theoretical template for the galaxy power spectrum, and is called galaxy bias \citep{2018PhR...733....1D,2000MNRAS.311..793B}.
%Galaxy bias can cause a systematic error in various cosmological analysis.

Therefore, 
this paper aims
to investigate the impact of the selection effect on 
the redshift-space power spectrum. 
To achieve this, 
%in this paper, 
we consider a case where an assumed sample of galaxies follows a redshift distribution, denoted as 
$n(z)$ or $n(\chi)$, 
representing the number density of galaxies per unit redshift interval or per unit radial comoving distance interval. 
We do not consider the redshift evolution of galaxy clustering (structure formation) to focus on the impact of the redshift-dependent selection effect. 
Hence in our setting, the {\em constant} comoving number density, $n(\chi)\mathrm{d}V(\chi)={\rm const.}$, corresponds to a 
homogeneous sample or the absence of the selection effect. 
We first use the linear theory to show that the selection function generally causes a bias in the multipole moments of the redshift-space galaxy power spectrum, 
along the Feldman-Kaiser-Peacock estimator \cite{FKP}. 
We then use halo catalogs from cosmological N-body simulations, \texttt{AbacusSumitt}~\citep{2021MNRAS.508.4017M,2021MNRAS.508..575G}, 
to construct mock catalogs of galaxies (halos), by selecting halos based on a redshift-dependent mass threshold 
such that the resulting redshift distribution of the selected halos follows the target 
$n(z)$ for the SDSS BOSS LOWZ- or CMASS-like galaxies.
We measure the multipole moments of the redshift-space power spectrum from the mock catalogs and then compare the moments with those obtained 
from halo catalogs of a single mass threshold, i.e. single tracers, in the same simulations.
This comparison quantifies the impact of the selection effect on the redshift-space power spectrum.

%possible systematic errors in existing cosmological parameter estimations based on galaxy clustering analysis such as BOSS \cite{2001AJ....122.2267E}, due to the inhomogeneity of the galaxy species in a observed sample. 
%For instance, if the galaxy sample is a flux-limited one, the species composing it are inhomogeneous along the line-of-sight.
%This is because we can detect only bright galaxies in the distant area from us, while we can detect also faint galaxies in the near area from us.
%On the other hand, the theoretical templates that are compared with the observed galaxy power spectrum always assume that the galaxy bias is homogeneous everywhere.
%This contradiction can cause a systematic error in cosmological analysis, and we call the undesired effect caused by our galaxy selection as \textit{selection effect} in this paper.
%So the topic we will address is the impact of the ignorance of the galaxy selection effect on the theoretical galaxy power spectrum prediction and cosmological parameter estimation. 

The structure of this paper is as follows:
In Section~\ref{sec:Methods} we give the details of the halo catalogs used, and describe our methods for generating galaxy mock catalogs as well as for measuring the redshift-space power spectrum from each mock catalog.
We also describe how we evaluate the galaxy selection effect.
Section~\ref{sec:results} gives the main results of this paper.
We evaluate the selection effect for several types of samples, and investigate a possible bias in the cosmological parameter estimation due to the ignorance of the selection effect.
In Section~\ref{sec:Conclusion} we give conclusion and discussion.

\section{Methods}\label{sec:Methods}
In this section, we describe the methodology used in this paper. 

\subsection{N-body Simulations and Halo Catalogs}\label{sub:halo_catalog}

%In this paper, 
We use the halo catalogs constructed from N-body simulation data for the {\it Planck}~2018 $\Lambda$CDM model, 
provided by \texttt{AbacusSummit}~\citep{2021MNRAS.508.4017M,2021MNRAS.508..575G}. 
\footnote{\url{https://abacussummit.readthedocs.io/en/latest/citation.html}}
The N-body simulations were performed using $6912^3$ particles in a comoving cubic box with 
side length of $2h^{-1}\mathrm{Gpc}$. The particle mass was $2\times 10^{9} h^{-1}\mathrm{M_{\odot}}$ and the force softening scale was $7.2h^{-1}\mathrm{kpc}$. In this paper we use only the halo catalog and do not use the N-body %simulations 
particle
data.
In particular, we use the halo catalogs at a single redshift output, 
$z=0.5$, 
because we focus on the impact of selection effect on clustering 
measurements and, in other words, we do not include the effect of redshift evolution in order to highlight our results.
Halos in each simulation realization were identified using \texttt{CompaSO} halo finder \citep{2022MNRAS.509..501H}. 
In this paper we use only central halos (labeled ``L0'' in the released catalogs). We use, as the position of each halo, the center-of-mass of the member particles 
and, as the velocity of each halo, the center-of-mass velocity for each halo, as provided in the halo catalogs.
We use 25 realizations of the halo catalogs 
%for the {\it Planck}~2018 cosmology  
to reduce the statistical errors in our results.

\subsection{Mock Galaxy Catalogs}
\label{sub:mocks_all}

We construct two types of mock galaxy catalogs. One of them includes the selection effect, and the other one is free from the effect. 
We will measure power spectrum from each mock catalog and take the ratio of the two power spectra to quantify the impact of the selection effect.
In the following, we describe how to create these two types of mock catalogs from the halo catalogs.
% in N-body simulations.

\subsubsection{Mock catalogs with galaxy selection effect}
\label{ssub:mocks_w-selectioneffect}

\begin{figure}[htbp]
\centering 
\includegraphics[width=1.0\textwidth]{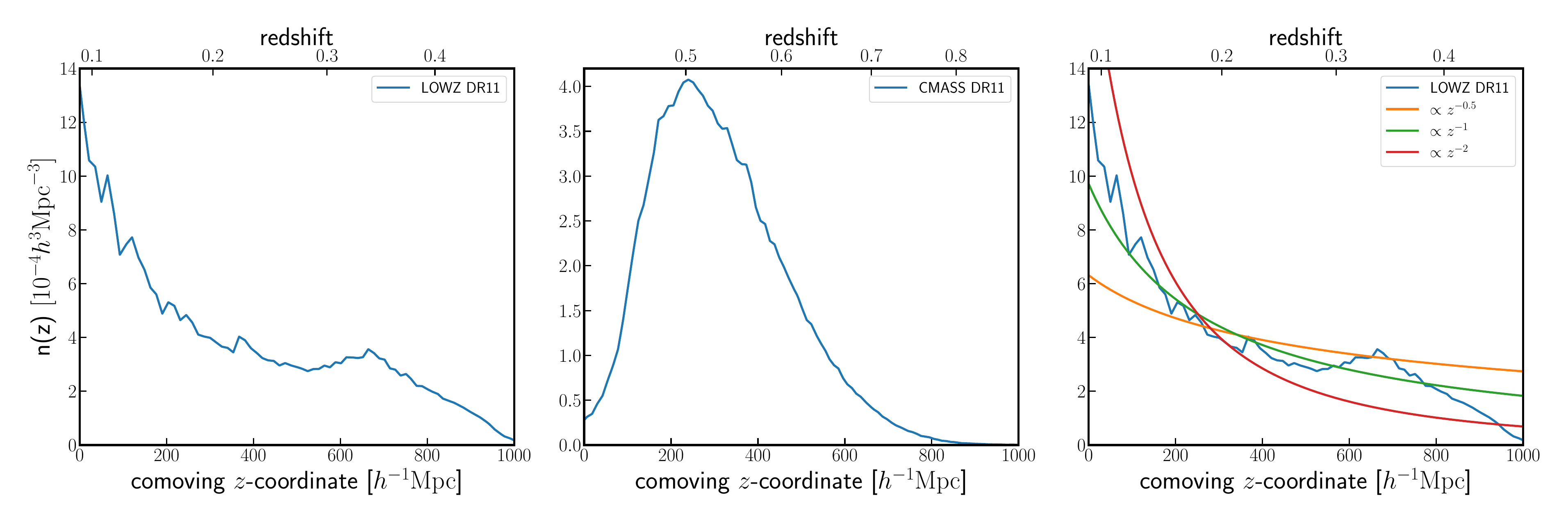}
\caption{\label{fig:SDSSdens} The number density of
LOWZ ({\em left panel}) and CMASS ({\em middle panel}) galaxies as a function of redshift, denoted as $n(z)$, taken from the SDSS DR11 catalog.
Note that $n(z)$ is the number density per unit redshift interval.
For illustrative purpose we also show the comoving $z$-coordinate on the bottom axis, computed using the inverse relation between 
redshift and the comoving radial distance at each redshift, $z=z^{-1}(\chi)$, based on the {\it Planck}~2018 cosmology. This $z$-coordinate can be compared to the size of \texttt{AbacusSumitt}
simulation box ($2~h^{-1}$Gpc) used in this paper.
The LOWZ galaxies are taken from the redshift range $z\simeq[0.09,0.5]$, while the CMASS galaxies are from the range $z\simeq [0.4,0.8]$. 
Note that we set the
comoving $z$-coordinate
at the lowest redshift to $\chi=0$ for each galaxy sample. 
We also consider a case of $n(z)$ following a power-law scaling given as $n(z)\propto z^{-\alpha}$ ({\em right panel}) to study the impact of galaxy selection effect as a general case.
}
\end{figure}

To mimic actual galaxy surveys, we consider the LOWZ and CMASS galaxy samples of Sloan Digital Sky Survey Baryon Oscillation Spectroscopic Survey (BOSS) Data Release 11 (DR11)\footnote{\url{https://data.sdss.org/sas/dr11/boss/lss/}}; 
we consider LOWZ galaxies at $z>0.09$,
and CMASS galaxies at $z>0.40$,
respectively, as shown in Fig.~\ref{fig:SDSSdens}.
In addition to these cases we also consider a selection function of galaxies that follows a power-law selection of $n(z)$, given by $n(z)\propto z^{-\alpha}$, for the generality of our discussion. 
We investigate three cases, $\alpha=0.5,1.0,2.0$,
and determine their normalizations so that the resulting averaged number density in the radial distance range of 
$[0,1000]h^{-1}$Mpc, which roughly corresponds to the range of LOWZ, becomes 
the same as in the LOWZ case (see the \textit{right panel} in Fig.~\ref{fig:SDSSdens}).
Throughout this paper we assume a {\it distant observer approximation}, and take the $z$-axis of each simulation box  to be along the line-of-sight direction.

\begin{figure}[htbp]
\includegraphics[width=1.0\textwidth]{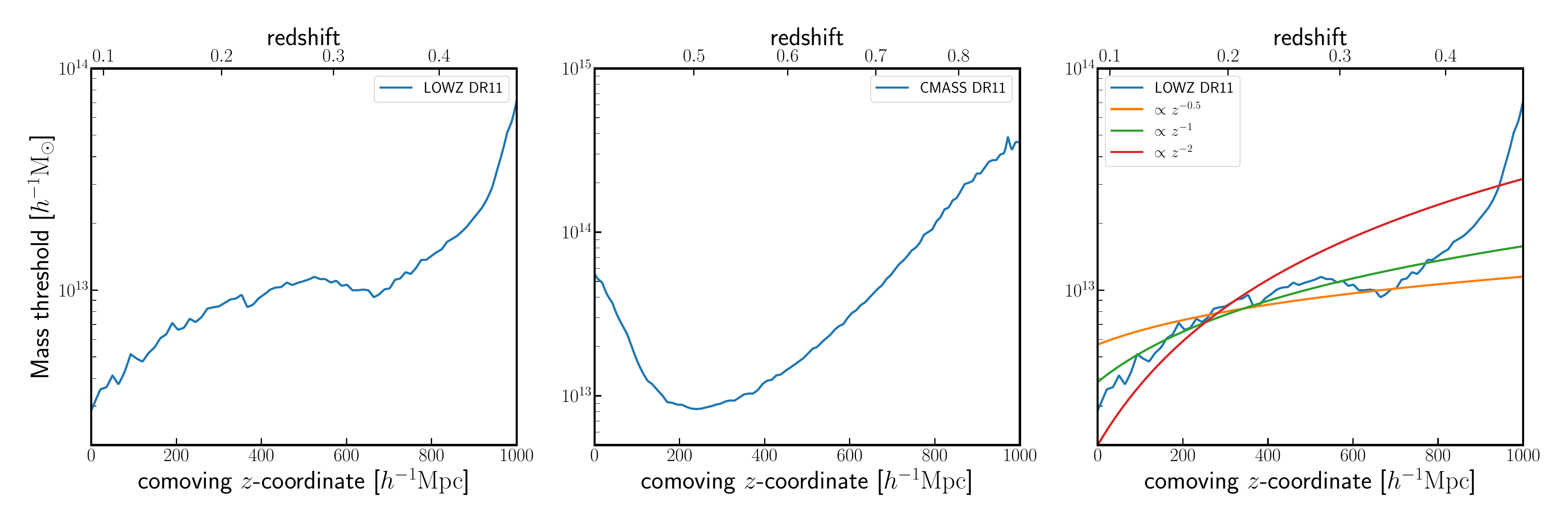}
\caption{\label{fig:Mmin_list} Halo 
mass thresholds as a function of redshift, $M_{\rm th}(z)$, for the LOWZ 
({\em left panel}) and 
CMASS ({\em middle panel}) samples, respectively. We select halos with mass above $M_{\rm th}(z)$ in each redshift bin, 
so that the resulting number density of halos reproduces the number density of each sample, $n(z)$, 
in Fig.~\ref{fig:SDSSdens} (also see Section~\ref{ssub:mocks_w-selectioneffect}). 
We determine the mass threshold function based on the average of 25 realizations of the halo catalog.
The three solid lines in the right panel show the mass threshold for each of the three power-law density 
cases in Fig.~\ref{fig:SDSSdens}.}
\end{figure}

The LOWZ and CMASS galaxies are selected based on the color and magnitude cuts \citep{2001AJ....122.2267E,2006MNRAS.372..425C,2009MNRAS.394L.107M,2016MNRAS.455.1553R}. It is very difficult and nearly
impossible to exactly mimic the actual color-magnitude cut for the galaxy selection using the halo catalog,
because it requires simulating galaxies while accounting for 
%detailed knowledge of 
the complex physics of galaxy formation and evolution. 
In this paper, we employ the following simplified selection function as a working example. 
For a given $n(z)$ for the sample of galaxies considered, we select halos with masses above the mass threshold $M_{\rm th}(z)$ 
in each simulation realization
so that the resulting number density matches the $n(z)$ in each redshift bin $[\chi,\chi+\mathrm{d}\chi]$ (since the comoving radial distance is given by $\chi=\chi(z)$).
We determine the mass threshold $M_{\rm th}(z)$ in each redshift bin from the average of the mass functions from 25 realizations of halo catalog, to minimize the sample variance effect.
Since heavier halos tend to host more massive and therefore brighter galaxies \citep{2018ARA&A..56..435W}, one could consider that 
this mass-threshold selection at each redshift mimics to some extent a flux limited sample of galaxies. 
In this paper, we refer to this method as \textit{abundance matching} (AM) method, and denote the power spectrum measured from this catalog as $P_{\rm AM}(k)$.
Fig.~\ref{fig:Mmin_list} shows the mass thresholds as a function of redshift,  $M_{\rm th}(z)$, which we use to construct mock catalogs for LOWZ, CMASS and power-law samples, respectively. 

We construct each mock galaxy catalog in a rectangular-shape region of
$2^2\times 1~(h^{-1}{\rm Gpc})^3$ volume, where we take the $z$-coordinate range of $[0,1000]~h^{-1}$Mpc from each simulation box
with a volume of $2^3~(h^{-1}{\rm Gpc})^3$, 
following Fig.~\ref{fig:SDSSdens}.
Thus we can
generate two mock catalogs from
each N-body simulation realization.
We apply zero padding to the halo distribution to perform the Fast Fourier Transform in a $2^3 ~(h^{-1}{\rm Gpc})^3$ cubic box
and 
then estimate the power spectrum using the Feldman-Kaiser-Peacock (FKP) estimator as described in Section~\ref{sec:fkp} in detail. 
Since the mock catalog 
%halo distribution 
violates the periodic boundary condition, the estimated power spectrum  
is affected by the window convolution. 
For the following results, we use 50 realizations of the mock catalogs for each galaxy sample. 

Halos of different masses have different clustering properties; heavier halos show greater clustering amplitudes and stronger nonlinearities, but have a fewer number density \citep{2019ApJ...884...29N}. 
Thus, the power spectrum measured from the above mock catalogs arises from superposition of power spectra between halos of different masses.

\subsubsection{Mock catalogs with a 
single halo-mass threshold}
\label{ssub:mock_single_mass_threshold}

%The mock catalogs we constructed above contain halos of different masses in each realization. 

To quantify the galaxy selection effect, we make a mock catalog without galaxy selection effect to be compared with the AM catalog. 
%
%We create such a galaxy catalog as follows:
We create such a galaxy catalog using
%in 
the following process:
%, which is common to all three galaxy samples:}
%the types of galaxy sample.}
\begin{itemize}
\item[(i)]
In each simulation, from halos with masses above a given mass threshold,
we randomly select halos in each redshift bin (or along the $z$-axis direction in our setting) so that the resulting redshift distribution of the number density follows $n(z)\propto \bar{n}_{\rm g}(z)$, where $\bar{n}_{\rm g}(z)$ is the redshift distribution for the LOWZ, CMASS or power-law galaxy sample, respectively, as shown in Fig.~\ref{fig:SDSSdens}.
When the number density of halos with masses above a given mass threshold is lower than the maximum number density of galaxies for each galaxy sample
at a certain redshift, we down-sample the halos while maintaining the shape of $n(z)$.
We estimate the power spectrum of the selected halos
using the FKP estimator (see below), which
retains the original power spectrum of the single mass threshold, and also 
incorporates the similar survey window effect as in the $P_{\rm AM}(k)$.
From 25 realization, we obtain 50 halo catalogs (2 catalogs from each realization as described above) and their spectra. 
%We prepare different catalogs of halos selected based on varying mass thresholds in each realization. 
%We create 50 halo catalogs for each mass threahold (2 catalogs from each realization as we described above). Note that, since we use the halo catalogs at a single redshift output ($z=0.5$), the avereage number density remains constant, i.e. has no dependecne on the $z$-axis. 
%prepare a large set of mass thresholds $M_{\rm th}$. We create $25$ halo catalogs for every mass threshold in the set.

%We construct different halo catalogs as a function of varying mass thresholds $M_{\rm th}$ in each realization.
\item[(ii)] 
Repeating the step (i), we find a certain mass threshold such that its (\textit{real-space}) galaxy power spectrum obtained in step (i) matches the amplitude of the target spectrum $P_{\rm AM}(k)$ on linear scales in the range $k=[0.02,0.1]~h{\rm Mpc}^{-1}$.
In this step we perform the $\chi^2$ fitting taking into account the sample variance error in each $k$ bin estimated from the variance of the 50 realizations. 
The halo catalog with the single mass threshold that minimizes the $\chi^2$ value is the catalog without a selection function, which we will use in the following analysis.
%For each halo catalog of a given mass threshold, we randomly select halos in each redshift bin (or along the $z$-axis direction in our setting) so that the resulting redshift distribution of the number density 
%satisfies follows $n(z)\propto \bar{n}_{\rm g}(z)$, where $\bar{n}_{\rm g}(z)$ is the redshift distribution for the LOWZ, CMASS or power-law galaxy sample, respectively, as shown in Fig.~\ref{fig:SDSSdens}.
%\reviewed{The reason that we select halos according to $\bar{n}(z)\propto \bar{n}_{\rm g}(z)$ is that 
%the numbe density of halos is lower than the maximum number density of galaxies for each galaxy sample
%at a certain redshift. }
%we show below.
% for the example. 
%The power spectrum estimated using the FKP estimator (see below)
%retains the original power spectrum of the single mass threshold, and also 
%incorporates the similar survey window effect as in the $P_{\rm AM}(k)$.
%for each of the three galaxy samples. 
%\reviewed{We get $50$ mock galaxy catalogs for every mass threshold.}
%\item[(iii)] We find a certain mass threshold such that its (\textit{real space}) galaxy power spectrum matches the target spectrum $P_{\rm AM}(k)$ 
%on the linear scales in the range $k=[0.02,0.1]~h{\rm Mpc}^{-1}$.
%In this step we perform the $\chi^2$ fitting taking into account the sample variance error in each $k$ bin estimated from the variance of the 50 realizations. 
%\reviewed{The halo catalog with a single mass threshold that minimizes the $\chi^2$ value is the catalog without a selection function, which we will use in the following analysis.}
\end{itemize}        
In the following we call this halo catalog as a {\em single mass-threshold} sample.
We will compare the two power spectra measured from the AM sample and the single mass-threshold sample to quantify the impact of the galaxy selection effect. 

\begin{figure}[htbp]
\centering 
\includegraphics[width=1.0\textwidth]{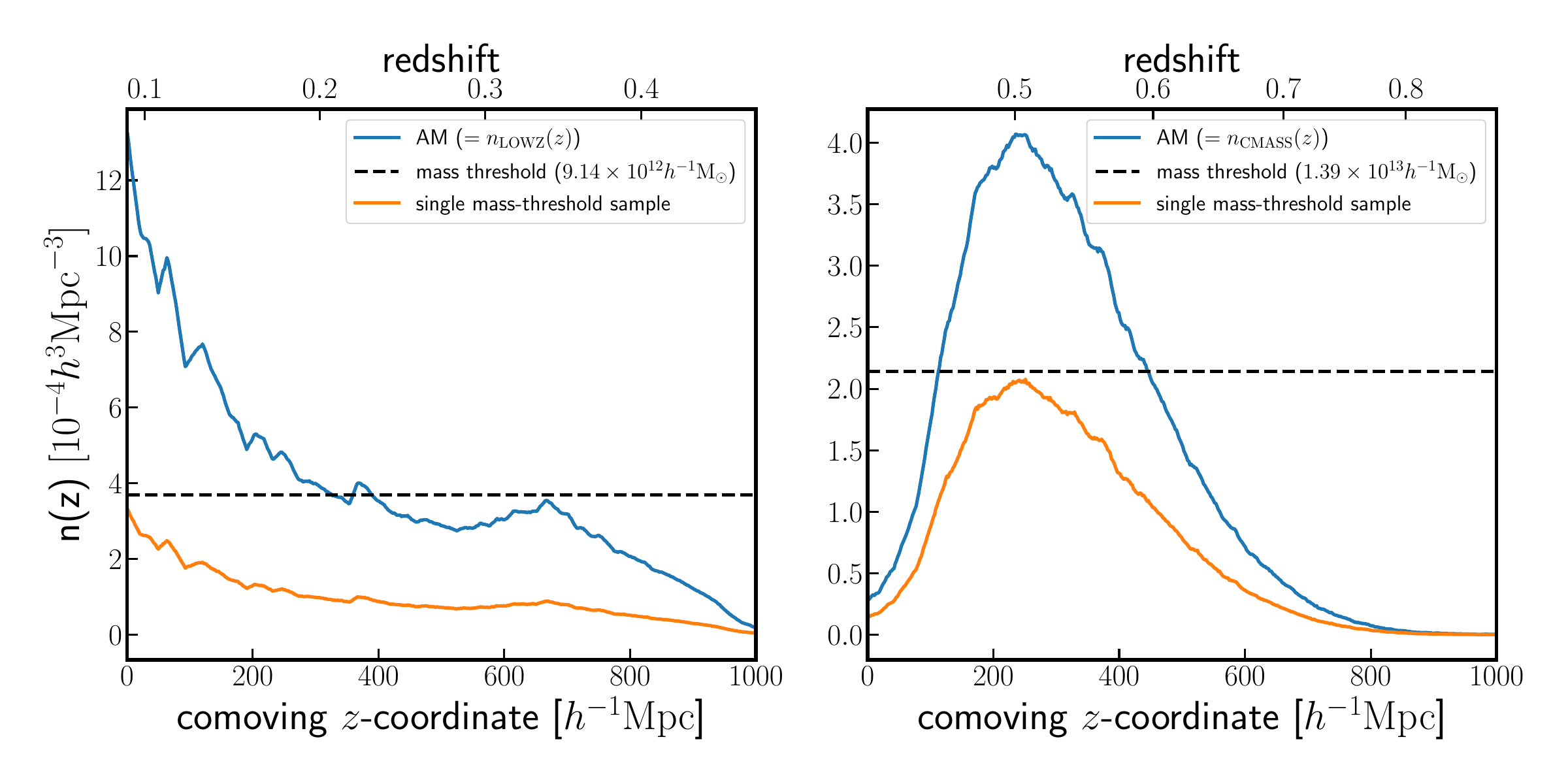}
\caption{A figure illustrating the method for incorporating the window effect on the single-mass threshold sample, used to quantify the impact of selection effects in comparison to the results of the AM catalog.
In each panel, the dashed line shows the number density which corresponds to the mass threshold in the legend.
For the single mass-threshold sample, 
we randomly select halos above this single mass threshold in each redshift bin 
such that the redshift dependence of the number density 
is proportional to that for the LOWZ, CMASS or power-law sample in Fig.~\ref{fig:SDSSdens} (here we show results only the 
for the LOWZ and CMASS samples) (see Section~\ref{ssub:mock_single_mass_threshold} for details on the single mass-threshold sample). 
This single mass-threshold hold sample has the same window function as that for the AM sample. 
\label{fig:window_convolution}
}
\end{figure}

Orange line in each panel of Fig.~\ref{fig:window_convolution} shows $n(z)$ of a single mass-threshold sample, constructed based on the above method, for the LOWZ or CMASS galaxy sample. 
The dashed line in each panel shows the number density of halos corresponding to a certain mass threshold, which by definition has a constant amplitude in the $z$ direction because we are using the halo catalogs 
at the single redshift output ($z=0.5$). 
Then, we randomly select halos in each radial distance bin  to satisfy
$n(z)\propto n_{\rm LOWZ}(z)$ or $n_{\rm CMASS}(z)$, respectively.
The proportional coefficient is determined by the ratio of the number density of halos of the single mass threshold (the dashed line in Fig.~\ref{fig:window_convolution}) to the peak number density 
of galaxies for each galaxy sample, $n_{\rm g}(z_{\rm peak})$; 
the number density at $z_{\rm peak}=0.08$ or $z_{\rm peak}\simeq 0.5$ for 
the LOWZ or CMASS sample, respectively.
%The single mass-threshold sample in Fig.~\ref{fig:window_convolution} has the power spectrum with the closest amplitude to the spectrum from the AM sample on the linear scales (see the above step \reviewed{(ii)}), 
%for each of the galaxy samples. 
For LOWZ sample, the  single mass threshold is $M_{\rm th}=9.14\times 10^{12}h^{-1}{\rm M_{\odot}}$, and the 
proportional factor in $n(z)\propto n_{\rm g}(z)$
is about $1/4$. For CMASS sample, the single mass-threshold is $M_{\rm th}=1.39\times 10^{12}h^{-1}{\rm M_{\odot}}$, and the proportional factor 
is about $1/2$.
The power spectrum measured from this single mass-threshold sample includes the survey window effect similarly to that for the AM galaxy sample
(see Section~\ref{sec:fkp}).
Note that the shot noise contamination estimated by the FKP estimator is different between the AM and single mass-threshold samples, and we  subtract it from the measured power spectrum.

In the following we denote the power spectrum measured from the single mass-threshold sample as $P_{M_{\rm th}}(k)$.

\subsection{FKP estimator}
\label{sec:fkp}

We use the Feldman-Kaiser-Peacock (FKP) estimator \cite{FKP} to measure power spectrum from each mock catalog. 
We define the galaxy number density field $F({\bf x})$ as
\begin{equation}
F(\mathbf{x}) = \frac{w(\mathbf{x})}{A_{\rm norm}^{1/2}}[n_{\rm g}(\mathbf{x}) - \alpha n_{\rm s}(\mathbf{x})] 
\label{Fx},
\end{equation}
where $n_{\rm g}(\mathbf{x})$ and $n_{\rm s}(\mathbf{x})$ are the 
number density fields for galaxies and randoms, respectively. 
$\alpha$ is a factor that adjusts the local mean number density of randoms to that of galaxies, defined as $\alpha = N_{\rm g}/N_{\rm s}$, where $N_{\rm g}$ is the total number of galaxy particles and $N_{\rm s}$ is that of random particles.
Throughout this paper, we adopt $\alpha=1/50$.
The normalization factor is $A_{\rm norm} = \int\!\mathrm{d}^3{\bf x}~ \bar{n}^2(\mathbf{x})w^2(\mathbf{x})$, and $w(\mathbf{x})$ is 
the so-called FKP weight to improve the signal-to-noise ratio in the power spectrum measurement in the shot noise regime:
\begin{equation}
w(\mathbf{x}) = \frac{1}{1+\bar{n}(\mathbf{x})P_0}, \label{FKP_weight}
\end{equation}
where $\bar{n}(\mathbf{x})$ is given by the redshift distribution $\bar{n}_{\rm g}(z)$
(Fig.~\ref{fig:SDSSdens}) (our mocks do not have any variations in the mean number density in the $x,y$-directions), and we adopt $P_0=10^4~h^{-3}\rm{Mpc}^{3}$ for 
the AM sample. 
For 
the single mass-threshold sample, we set $P_0=10^4/\beta~h^{-3}\rm{Mpc}^{3}$ in Eq.~(\ref{FKP_weight}), 
where the $\beta$ is the proportional factor of $\bar{n}_{M_{\rm th}}(z)\propto\bar{n}_{\rm g}(z)$
in Section~\ref{ssub:mock_single_mass_threshold}. 
This scaling by $\beta$ is necessary to keep the \textit{window function} (see Eq.~\ref{expectatoin_value}) the same between AM sample and single mass-threshold sample. 
We generate the number density field of random particles $n_s(\mathbf{x})$ by repeating a Bernoulli process whose probability of success has the redshift dependence proportional to $\bar{n}_g(z)$, while the particles are randomly distributed in the $x$, $y$ directions.

We then perform the fast Fourier transform of $F(\mathbf{x})$ and estimate the power spectrum from
\begin{equation}
\hat{P}(\mathbf{k}) \equiv |F(\mathbf{k})|^2 - P_{\rm sn}\label{P_est},
\end{equation}
where $P_{\rm sn}$ is the shot noise contamination, estimated by
\begin{align}
P_{\rm sn}&=\frac{(1+\alpha)\int \mathrm{d}^3{\bf x}~ \bar{n}(\mathbf{x})w^2(\mathbf{x})}
{\int \mathrm{d}^3{\bf x}~ \bar{n}^2(\mathbf{x})w^2(\mathbf{x})}.
\label{eq:Psn_def}
\end{align}
We use $1024^3$ grids in a cubic volume with a side length of $2~h^{-1}{\rm Gpc}$ that covers the entire region of each mock catalog. 
Note that we use zero padding in grids where the data does not exist.
In this paper, we employ Cloud In Cell (CIC) assignment scheme to generate grid-based data from the particle (galaxy and random) data. 
We correct for the CIC kernel effect when measuring the power spectrum, which is significant on small scales relevant to the grid scale \citep{2005ApJ...620..559J}. 
The FKP method gives an estimate of the underlying power spectrum including the convolution of the survey window effect:
\begin{align}
\braket{\hat{P}(\mathbf{k})}&=\braket{|F(\mathbf{k})|^2} -P_{\rm sn} \notag \\
&= \int \frac{\mathrm{d}^3{\bf k}^{'}}{(2\pi)^3} P(\mathbf{k^{'}})|G(\mathbf{k}-\mathbf{k^{'}})|^2 
\label{expectatoin_value},
\end{align}
where $G(\mathbf{k})\equiv[{\int \mathrm{d}^3{\bf x}~ \bar{n}(\mathbf{x})w(\mathbf{x})e^{i\mathbf{k}\cdot\mathbf{x}}}]/[{[\int \mathrm{d}^3{\bf x}~ \bar{n}^2(\mathbf{x})w^2(\mathbf{x})]^{1/2}}]$ is a \textit{window function} \citep{2013IJAA....3..243S,Beutler_2014}. 
For $|{\bf k}|\gg 1/L$, where $L$ is a size of a survey window, $\langle \hat{P}(k)\rangle\simeq P(k)$ \citep[e.g.,][]{2013PhRvD..87l3504T}, i.e. the estimator gives an unbiased estimate of the underlying power spectrum. 
Since the AM mocks and the single mass-threshold mocks have 
the same redshift dependence of halo distribution, $n_{\rm AM}(z)\propto n_{M_{\rm th}}(z)$, by design, 
the power spectra measured from these two mocks have the same effect of window convolution as long as we keep their $w(\mathbf{x})$ the same.

The normalization factor $A_{\rm norm}$ in Eq.~(\ref{Fx}) and the shot noise term $P_{\rm sn}$ in Eq.~(\ref{eq:Psn_def}) are given by
\begin{align}
&A_{\rm norm} \to \alpha \sum_{i=1}^{N_s}\bar{n}(z_i)w(z_i)^2 ,\\
P_{\rm sn} &\to \frac{1}{A_{\rm norm}} \Big(\sum_{i=1}^{N_g} + \alpha^2 \sum_{i=1}^{N_s} \Big)w(z_i)^2. \label{P_shot_est}
\end{align}
We also use the interlacing scheme in Ref.~\cite{GridAssign} to mitigate the aliasing effect.
We show the results measured from 
50~realizations of the mock catalog for each galaxy sample.

\subsubsection{Linear limit of $P_{\rm AM}$}
\label{sec:linear_limit}

As we discussed, in an actual galaxy survey, it is impossible to have a homogeneous sample of the underlying halos, and rather we have an inhomogeneous sample  of halos with respect to halo masses along the radial direction, i.e. affected by the selection effect. This means that the underlying power spectrum becomes position-dependent: 
$P({\bf k};{\bf x})$.
We can evaluate
the expectation value of Eq.~(\ref{P_est}) for the AM mock catalog as
\begin{align}
\braket{\hat{P}_{\rm AM}(\mathbf{k})}=\frac{\int \mathrm{d}^3{\bf x} ~ 
\bar{n}^2(\mathbf{x})w^2(\mathbf{x})P_{\rm hh}(\mathbf{k};M_{\rm th}(\mathbf{x}))}{\int \mathrm{d}^3{\bf x} ~ \bar{n}^2(\mathbf{x})w^2(\mathbf{x})}=\frac{\int \mathrm{d}x_3 ~ \bar{n}^2(x_3)w^2(x_3)P_{\rm hh}(\mathbf{k};M_{\rm th}(x_3))}
{\int \mathrm{d}x_3 ~ \bar{n}^2(x_3)w^2(x_3)}\label{powerspectrum_selection},
\end{align}
where we 
ignored the window effect for simplicity,
and we set the $x_3$ direction to be along the redshift ($z$) direction.
We have dropped the $x_1x_2$ integrals in the second equality of the above equation, since we employ a homogeneous window in the $x_1x_2$ plane under the distant observer approximation, and
$M_{\rm th}(x_3)$ is the halo mass threshold in each $x_3$ bin as shown in Fig.~\ref{fig:Mmin_list}.
Eq.~(\ref{powerspectrum_selection}) shows that, in the presence of the selection effect, the power spectrum
should be given by the integral of
power spectra for halos of
different masses along the redshift direction. 

Before going to the results, it would be useful to consider the linear limit of Eq.~(\ref{powerspectrum_selection}). 
In the real-space case, we can express the power spectrum at the position $x_3$ 
as $P_{\rm hh}(k; M_{\rm th}(x_3))=b_1^2(M_{\rm th}(x_3))P_{\rm mm}^{\rm L}(k)$ in the linear regime, where 
$b_1(M_{\rm th})$ is the linear bias parameter of halos above mass threshold $M_{\rm th}$ and
$P^{\rm L}_{\rm mm}$ is the linear matter power spectrum.
Note that, since we use the single redshift output of simulation,
we have ignored the redshift dependence of the growth factor
here and in the following.
Then the real-space power spectrum measured by the FKP estimator from the AM samples 
in the linear regime is given as
\begin{align}
\braket{\hat{P}^{\rm R}_{\rm AM}(k)}\simeq \overline{(b_1)^2}P^L_{\rm mm}(k),
\label{eq:pkreal_linearlimit}
\end{align}
with
\begin{align}
\overline{(b_1)^2}\equiv \frac{1}{\int\!\mathrm{d}x_3~\bar{n}^2(x_3)w^2(x_3)}
\int\!\mathrm{d}x_3~ \bar{n}^2(x_3)w^2(x_3) [b_1\!\left(M_{\rm th}\!(x_3)\right)]^2.
\label{eq:ave_bh_squared}
\end{align}
Thus, the overall amplitude of $\hat{P}^{\rm R}_{\rm AM}(k)$ in the linear regime is proportional to the radial-distance average of 
$b_1^2(x_3)$ weighted by $\bar{n}^2(x_3)w^2(x_3)$.

Similarly, in redshift space, we can express
the local redshift-space power spectrum at position $x_3$
as $P_{\rm hh}^{\rm S}(\mathbf{k},M_{\rm th}(x_3))=[b_1^2(M_{\rm th}(x_3))+2f\mu^2 b_1(M_{\rm th}(x_3))+f^2\mu^4]P^{\rm L}_{\rm mm}(k)$ in the linear regime, based on the Kaiser formula \cite{1987MNRAS.227....1K}, where $f$ is the logarithmic growth rate and $\mu$ is the cosine angle between the line of sight direction 
($x_3)$ and the wavenumber.
Therefore the redshift-space power spectrum measured by the FKP estimator for the AM samples 
is 
\begin{align}
\braket{\hat{P}^{\rm S}_{\rm AM}(k,\mu)}\simeq \left[\overline{(b_1)^2}+2f\mu^2\overline{b_1}+f^2\mu^4\right]P^{\rm L}_{\rm mm}(k),
\end{align}
where
\begin{align}
\overline{b_1}\equiv \frac{1}{\int\!\mathrm{d}x_3~\bar{n}^2(x_3)w^2(x_3)}
\int\!\mathrm{d}x_3~ \bar{n}^2(x_3)w^2(x_3) b_1\!\left(M_{\rm th}\!(x_3)\right).
\label{eq:ave_bh}
\end{align}
It should be noted that $\overline{(b_1)^2}$ is generally different from $(\overline{b_1})^2$ for a sample of halos of different masses.
This means that, even if we define the single mass-threshold sample that reproduces the amplitude of 
the {\em real-space} power spectrum for the AM sample on linear scales, i.e. $P_{M_{\rm th}}^{\rm R}(k)\simeq P_{\rm AM}^{\rm R}(k)$, 
the redshift-space power spectra for the two samples 
do not necessarily match at small $k$ in the linear regime. 
Strictly speaking, therefore, the redshift-space power spectrum in the presence of the selection effect cannot generally be 
interpreted in terms of the power spectrum of {\em single} biased tracers as usually done in analyses.
In the following, we quantify the impact of the selection effect.

A similar argument would hold on nonlinear scales. On quasi nonlinear scales, one can use 
the perturbative bias expansion to model the galaxy density field, e.g. using the effective field theory of 
large-scale structure (EFTofLSS) \citep{2012JCAP...07..051B,2022arXiv221208488I,2020PhRvD.102l3541N,2018PhR...733....1D}.
The linear and higher-order bias parameters and the EFTofLSS counter terms are known to depend on halo mass or properties of galaxies \citep{2018PhR...733....1D,2024arXiv241201888I,2024arXiv241008998A}. This means that, when the selection effect given by $n(z)$ exists, it causes non-trivial redshift-dependent effects in the clustering properties, which  therefore cannot be described by single tracers. On very nonlinear, small scales, phenomenological approaches such the as halo occupation distribution (HOD) method \citep[e.g.][]{kobayashi2022fullshapecosmologyanalysissdssiii}
and subhalo abundance matching method \citep[e.g.][]{2006ApJ...647..201C} can be used to model clustering properties of galaxies. In these approaches, galaxies are populated in their host halos according to the given galaxy-halo connection, and therefore the selection effect inevitably leads to redshift-dependent effects in the clustering properties. 

From these considerations, we conclude that the selection effect causes redshift-dependent systematic effects in the clustering properties of galaxies that cannot be described by single tracers, on all scales from the linear to nonlinear scales. The question then arises of how large the systematic effects are, which is the main focus of this paper.

\section{Results}
\label{sec:results}

In this section, we show the main results of this paper. 

\subsection{The selection effect on real- and redshift-space power spectra for the LOWZ
and CMASS mock catalogs}
\begin{figure}[htbp]
\centering 
\includegraphics[width=1.0\textwidth]{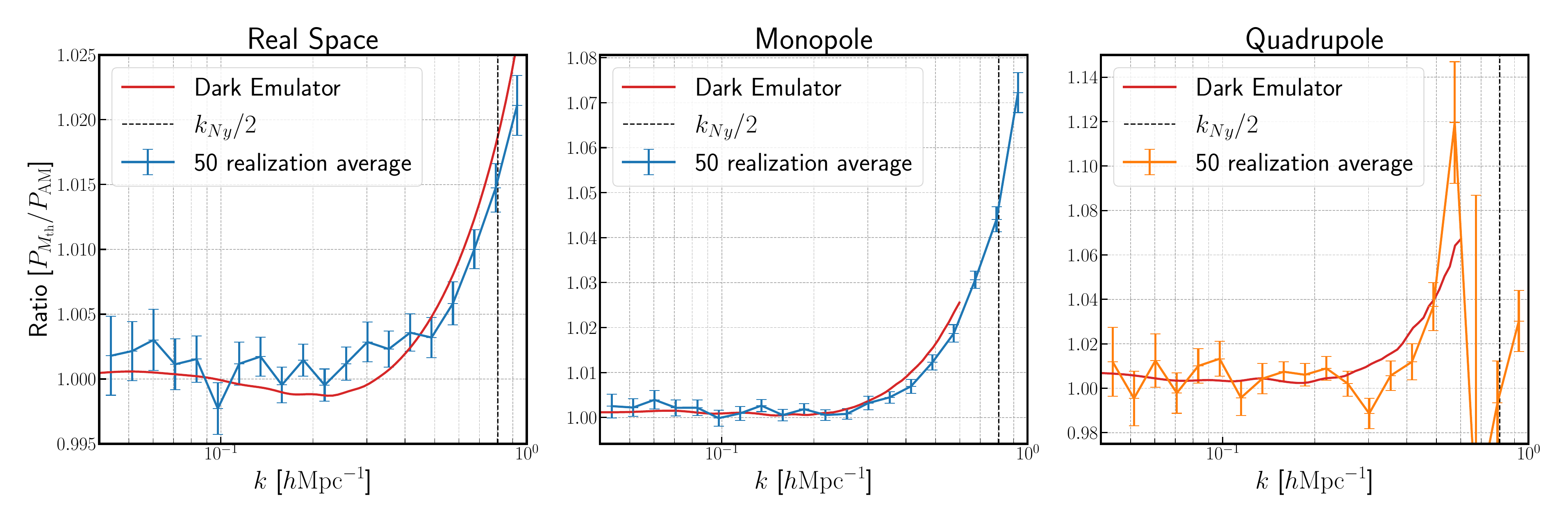}
\caption{\label{fig:results_LOWZ}
A comparison of the two power spectra, 
%The comparison of the two power spectra, 
$P_{\rm AM}$ and $P_{M_{\rm th}}$, for the abundance-matching (AM) sample 
and the single mass-threshold ($M_{\rm th}$) sample both of which reproduce $n(z)$ of the LOWZ sample. 
%for LOWZ samples.
We show the ratio for the real-space power spectrum ($\textit{left panel}$), and the monopole ($\textit{middle panel}$) and quadrupole ($\textit{right panel}$) moments of the redshift-space power spectrum. The data points in each panel are 
%These plots are 
the mean of the 
$50$ realizations and the error bars are the $1\sigma$ error on the mean, estimated by dividing the standard deviation by $\sqrt{50}$. 
The black dashed vertical line in each panel  indicates the half of the Nyquist wavenumber, and the solid line is 
the ratio computed using the {\texttt Dark~Emulator} in Kobayashi et al. \citep{kobayashi2022fullshapecosmologyanalysissdssiii} (see text for the details). 
}
\end{figure}
Fig.~\ref{fig:results_LOWZ} shows the results for the LOWZ galaxy sample. 
We can measure the ratio $P_{M_{\rm th}}/P_{\text{\rm AM}}$ from the 50 realizations.
We adopt
$M_{\rm th}=9.14\times 10^{12}h^{-1}\rm M_{\odot}$
to define the single mass-threshold sample, which minimizes the $\chi^2$ difference between 
the real-space power spectra $P_{\rm AM}$ and $P_{M_{\rm th}}$ in the range $k=[0.02:0.1] ~ h\rm Mpc^{-1}$ as shown in the left panel. 
The blue or orange points in each panel are the mean of the 50 realizations, and the error bars are the $1\sigma$ errors on the mean, 
estimated by dividing the standard deviation by $\sqrt{50}$.
The black dashed line indicates the half of the Nyquist wavenumber.

The \textit{left panel} of Fig.~\ref{fig:results_LOWZ} shows the ratio of the real space power spectra for %the two galaxy samples. 
the AM and the single mass-threshold mock catalogs. 
The ratio is very close to unity to within 0.5\% up to $k\simeq 0.5~h^{-1}{\rm Mpc}$, 
but starts to deviate from unity in the larger $k$ bins. 
The nice agreement also holds for the monopole and quadrupole moments of the redshift-space power spectrum 
in the middle and right panels, respectively. The rapid change around $k\sim 0.6~h{\rm Mpc}^{-1}$ for the quadrupole moment is due to the fact that the moment has a zero crossing around the scale.
Thus, we conclude that the selection effect is small for the LOWZ-like sample.

\begin{figure}[htbp]
\centering 
\includegraphics[width=1.0\textwidth]{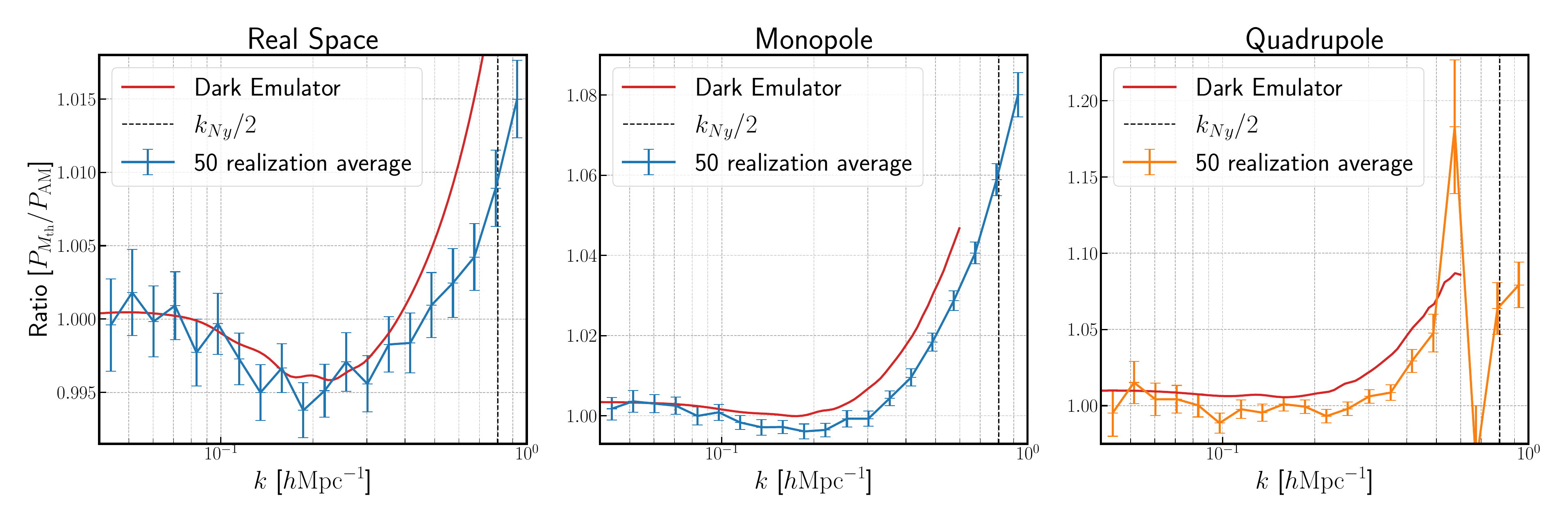}
\caption{\label{fig:results_CMASS}
Similar to the previous figure, but the results for the CMASS sample. 
}
\end{figure}
Fig.~\ref{fig:results_CMASS} shows the similar results for the CMASS galaxy sample. 
We adopt 
$M_{\rm th}=1.39\times 10^{13}h^{-1}\rm M_{\odot}$ for the single mass-threshold sample.  
As in Fig.~\ref{fig:results_LOWZ}, we find a similar degree of the agreement for the real- and redshift-space power spectra. 

We now give a theoretical interpretation of the results in Figs.~\ref{fig:results_LOWZ} and \ref{fig:results_CMASS}.
For this purpose, we use the \texttt{Dark~Emulator} developed in Ref.~\cite{2020PhRvD.102f3504K}
\citep[also see][]{2019ApJ...884...29N}, which is a simulation-based emulator that allows a fast, accurate computation 
of the real- or redshift-space power spectrum of halos for input parameters (cosmological parameters, redshift, and halo masses). For the redshift-space power spectrum, the emulator outputs $P(k,\mu)$, where 
$\mu$ is a cosine angle between the wavenumber vector and the line-of-sight direction. Hence, we can use
the emulator output to compute the monopole and quadrupole moments. 
The red solid curves in Figs.~\ref{fig:results_LOWZ} and \ref{fig:results_CMASS} show the ratio, $P_{M_{\rm th}}/P_{\rm AM}$, computed using \texttt{Dark~Emulator}.
We can compute $P_{\rm AM}$ based on Eq.~(\ref{powerspectrum_selection}), substituting the emulator's output for the local power spectrum,
\footnote{For the actual computation of the local power spectrum in Eq.~(\ref{powerspectrum_selection}), we input the cumulative halo number density $\bar{n}(x_3)$ (Fig.~\ref{fig:SDSSdens}) into the emulator, instead of halo mass threshold $M_{\rm th}(x_3)$ (Fig.~\ref{fig:Mmin_list}).}
while we can compute $P_{M_{\rm th}}$ directly using the emulator prediction for a single mass threshold.
We determine this mass threshold so that $P_{M_{\rm th}}$ matches $P_{\rm AM}$ at $k=0.02~h\rm Mpc^{-1}$.
These emulator predictions fairly well reproduce the simulation result, including the $k$ dependence. 
Thus, these results provide independent justification for the result that the selection function is small.
Please note that the emulator predictions for the monopole and quadrupole moments 
are about 1\% and 5\% fractional accuracies for halos of $\sim 10^{13}h^{-1}M_\odot$ up to $k\simeq 0.60~h{\rm Mpc}^{-1}$
for cosmologies around the fiducial $\Lambda$CDM model, as carefully studied in Ref.~\cite{2020PhRvD.102f3504K}.

We can also interpret the upturn behavior in high $k$ bins in 
Figs.~\ref{fig:results_LOWZ} and \ref{fig:results_CMASS} qualitatively as follows. The average mass of halos in the AM catalog 
is {\em lighter} than the halo mass in the single mass-threshold catalog for each realization.
The nonlinear clustering for halos of lower mass is weaker than that for heavier halos. For example,  
lighter halos have smaller nonlinear bias parameter ($b_2$) than heavier halos do \citep{HaloBias,2018PhR...733....1D}. 
This results in the amplitude of the power spectrum at higher $k$ for lighter halos being smaller than that for heavier halos, even when the linear-scale amplitudes of the power spectra are matched (see Eq.~\ref{eq:ave_bh_squared}).
In addition, lighter halos have larger random motions in the large-scale structure than heavier halos do, leading to a larger suppression in the redshift-space power spectrum amplitudes for lighter halos
via the Finger-of-God like effect as shown in Fig.~3 of
\citep{2011PhRvD..84d3526N}. 
These effects lead to the upturn behavior at high $k$ for the ratio of $P_{M_{\rm th}}/P_{\rm AM}$. 

\begin{figure}[htbp]
\centering 
\includegraphics[width=1.0\textwidth]{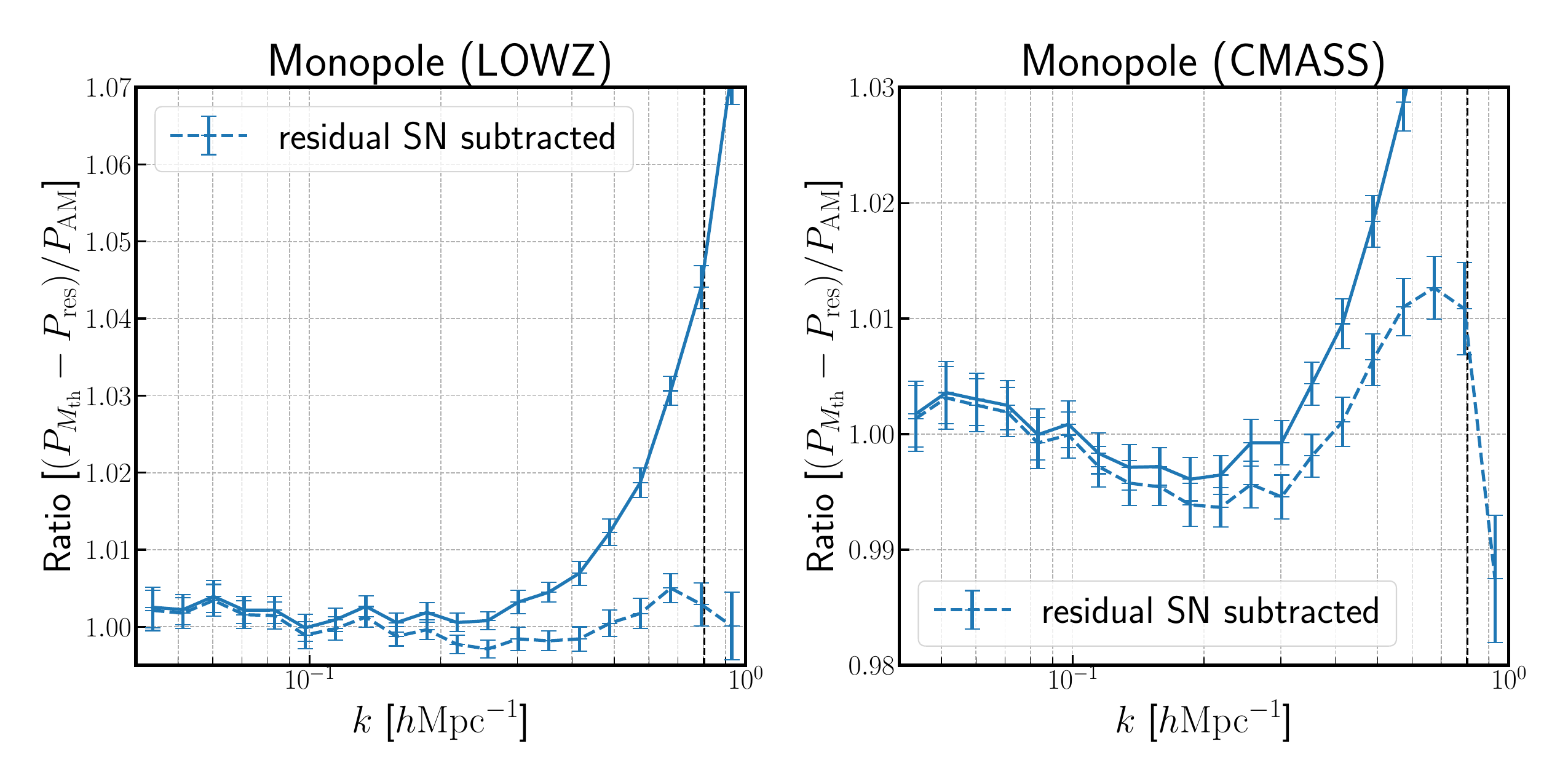}
\caption{\label{fig:residual_shotnoise}
Similarly to Figs.~\ref{fig:results_LOWZ} and \ref{fig:results_CMASS}, 
but here we subtract the residual shot noise from the measured power spectrum for $P_{M_{\rm th}}(k)$ before comparing it with 
$P_{\rm AM}(k)$.
%included the residual shot noise. 
We introduce the residual shot noise parameter as a free parameter for $P_{M_{\rm th}}(k)$
and determine its best-fit value by minimizing the difference between the shot-noise-subtracted 
$P_{M_{\rm th}}(k)$ and $P_{\rm AM}(k)$ over the range $k=[0.02,0.3]~h{\rm Mpc}^{-1}$.
The dashed line in each panel is for the ratio when using $P_{M_{\rm th}}(k)$ with the residual shot noise subtracted,  
while the solid line is the same as in Figs.~\ref{fig:results_LOWZ} and \ref{fig:results_CMASS}.
%The absorption of the uptrun behavior to the nuisance parameter $P_{\rm res}$.
%We confirm that the upturn behavior in monopole disappears within $\sim1.2\%$ if we subtract an appropriate constant $P_{\rm res}$ from $P_{M_{\rm th}}$.
}
\end{figure}

These upturn behaviors could also be attributed to the effect of residual shot noise
\cite{2013PhRvD..88h3507B}. 
In standard cosmological analyses of the power spectrum \citep[e.g.,][]{Ivanov_2020}, residual shot noise is 
commonly included as a nuisance parameter.
We determine the residual shot noise by minimizing the the difference between the shot-noise-subtracted 
$P_{M_{\rm th}}(k)$ and $P_{\rm AM}(k)$ over the range $k=[0.02,0.3]~h{\rm Mpc}^{-1}$. 
The residual shot noise estimated in this way is: $P_{\rm res}\sim 16$ or $20~(h{\rm Mpc}^{-1})^3$
for the LOWZ or CMASS sample, respectively, compared to the naive shot noise $1/\bar{n}_{\rm g}\sim 10^4~(h{\rm Mpc}^{-1})^3$. 
We confirm that the upturn behavior in the monopole is reduced if we include 
the residual shot noise parameter for $P_{\rm M_{\rm th}}$.
Note that the quadrupole, or more generally the higher-order moments, are not affected by the residual shot noise.
 
\subsection{
The results for halo samples with power-law $\bar{n}(z)$}
\label{sec:power-law}

\begin{figure}[htbp]
\centering 
\includegraphics[width=1.0\textwidth]{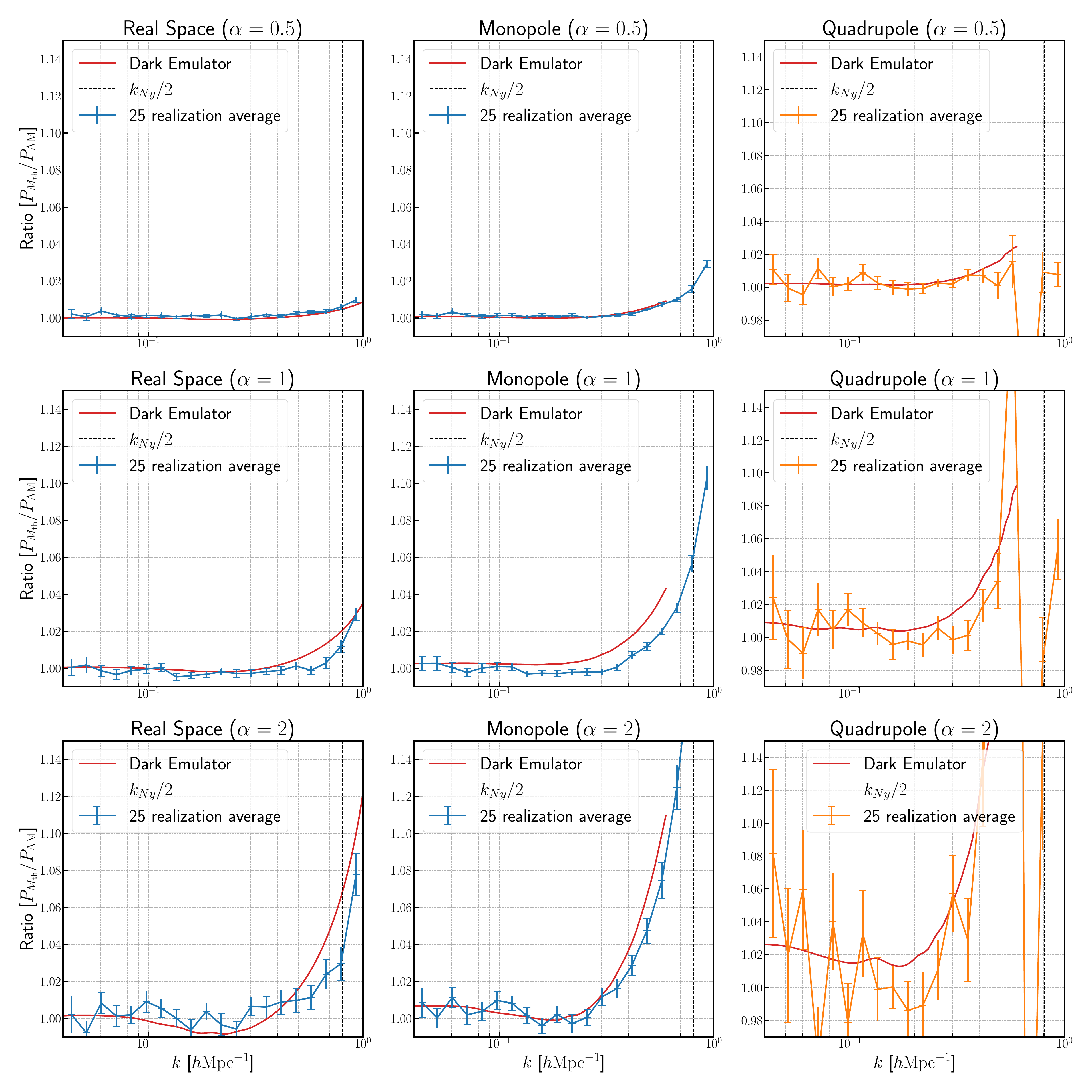}
\caption{\label{fig:power_law_ratio}Similar to Fig.~\ref{fig:results_LOWZ}, but the results 
for the power-law samples in the right panel of Fig.~\ref{fig:SDSSdens}. 
The panels in the top, middle and bottom row show the results for $\alpha=0.5, 1$ and $2$ of $n(z)\propto z^{-\alpha}$, 
respectively. 
%The comparison of the two power spectra, $P_{\rm AM}$ and $P_{M_{\rm th}}$, for power-law $z^{-\alpha}$ samples.We show the ratios for the real-space power spectrum ($\textit{left column}$), monopole ($\textit{middle column}$) and quadrupole ($\textit{right column}$). These plots are the mean value of the $25$ realizations and its error bars are their unbiased standard deviations divided by $\sqrt{25}$. The red lines are our theoretical explanation.
}
\end{figure}
To have the generality of our discussion, we also study the impact of the selection effect for a galaxy sample with 
$\bar{n}(z)\propto z^{-\alpha}$ ($\alpha=0.5, 1$ or $2$) as shown in the right panel of Fig.~\ref{fig:SDSSdens}.
%We follow nearly the same procedure in Sections~\ref{sub:mocks_all} and \ref{sec:fkp} to create the mock catalogs and measure the power spectra from them, but here we do not apply zero padding when creating mock catalogs in the $2^3 ~(h^{-1}{\rm Gpc})^3$ cubic box, since
%we have the halo distribution inside the whole original N-body box, unlike the LOWZ or CMASS samples.
Fig.~\ref{fig:power_law_ratio} shows the results.
For the cases of
$\alpha=0.5,1$, we find that the ratio is close to unity up to $k\simeq 0.2~h{\rm Mpc}^{-1}$, 
but starts to show an upturn behavior in high $k$ bins as in Figs.~\ref{fig:results_LOWZ} and \ref{fig:results_CMASS}.
We can also find that the emulator results (red curves) fairly well reproduce the simulation results.
However, for the case of $\alpha=2$, which corresponds to the largest selection effect in our study, 
the deviation from unity in the ratio can be significant:
%A closer look shows that the deviation from unity in the ratio is the largest for the case of $\alpha=2$ 
%as expected; 
the monopole moment shows a small deviation, up to 
%shows a deviation 
$1\%$ at $k\lesssim 0.3~h{\rm Mpc}^{-1}$,
%in the linear regime, 
whereas the quadrupole exhibits a large deviation of up to $8\%$,
due to the reason explained in Section~\ref{sec:linear_limit}.
This implies that if $n_{\rm g}(z)$ has a large redshift dependence,
the selection effect could be large and not negligible.

\subsection{The impact of the selection effect on cosmological parameter estimation}
\label{sec:estimation_bias}

\begin{figure}[htbp]
\centering 
\includegraphics[width=1.0\textwidth]{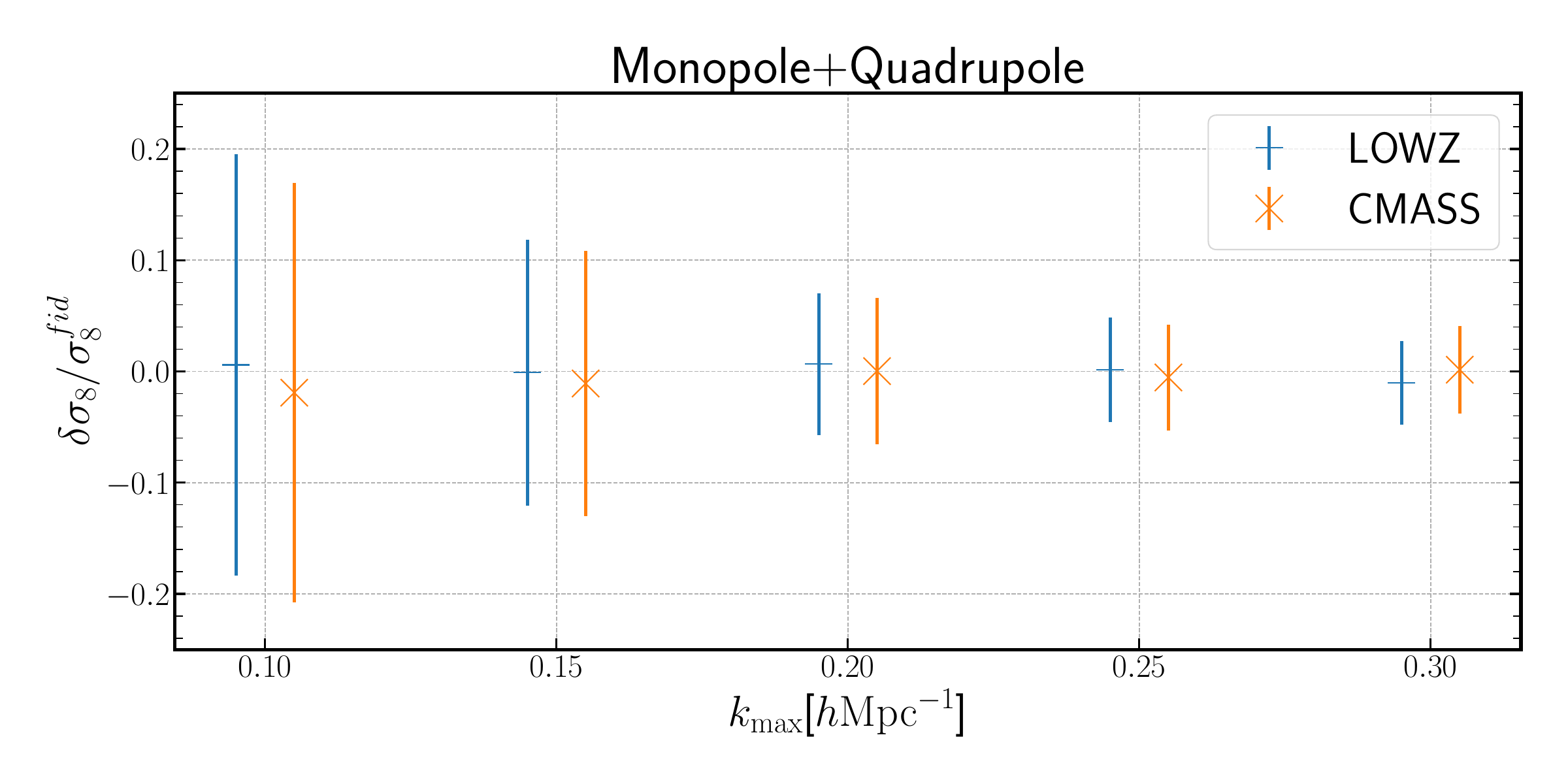}
\caption{\label{fig:bias_in_estimation}
%An estimation of a 
A possible bias in the $\sigma_8$ estimation from hypothetical measurements of the monopole and 
quadrupole moments of the redshift-space power spectrum for the LOWZ- and CMASS-like galaxy samples, if 
the selection effect is ignored. 
%due to the selection effect, if the effect is ignored in the model prediction of the redshift-space (halo-halo) power spectrum, $P_0$ and $P_2$. 
We estimate the bias using
%based on 
the Fisher method (Eq.~\ref{fisher_bias}) and present the bias as a function of the maximum wavenumber 
$k_{\rm max}$ included in the analysis. 
The error bar corresponds to a volume of $2~(h^{-1}{\rm Gpc})^3$ and includes marginalization over other parameters 
(see Section~\ref{sec:estimation_bias} for the details).
We slightly shift the symbols along the $x$-axis for illustration.
%We show the bias relative to the fiducial value (true value) of each parameter, as a function of the maximum wavenumber $k_{\rm max}$, and also show the marginalized error for comparison.
}
\end{figure}
In this section, we assess the impact of the selection effect on cosmological parameter estimation.
Since the selection effect causes a bias in the power spectrum amplitude and does not shift the scale of the baryon acoustic oscillations (BAO) as can be found from Figs.~\ref{fig:results_LOWZ}--\ref{fig:power_law_ratio}, 
we evaluate the impact of 
the selection effect on an estimation of $\sigma_8$, the present-day rms linear mass density fluctuations within a top-hat sphere of $8~h^{-1}{\rm Mpc}$ radius, which is
one of the most important parameters estimated from galaxy clustering analysis
\citep{2020PhRvD.101b3510K,kobayashi2022fullshapecosmologyanalysissdssiii}.
We use the Fisher matrix formalism \citep{2005APh....23..369H,2020PhRvD.101b3510K} to estimate a bias in cosmological parameters as
\begin{align}
\delta p_{\alpha} &= \sum_{\ell \ell^{'}}\sum_{\beta} \Big( F^{\ell \ell^{'}} \Big)^{-1}_{\alpha \beta} \sum_{ij} \Big[ P_{\ell}^{\rm selection}(k_i) - P_{\ell}^{\rm w/o~selection}(k_i) \Big] {\rm Cov}^{-1} \Big[ \hat{P}_{\ell}(k_i), \hat{P}_{\ell^{'}}(k_j) \Big] \frac{\partial P_{\ell^{'}}(k_j)}{\partial p_{\beta}} \notag \\
&\simeq \sum_{\ell \ell'}\sum_{\beta} \Big( F^{\ell \ell^{'}} \Big)^{-1}_{\alpha \beta} \sum_{ij} \Big( \frac{P_{{\rm AM},\ell}(k_i)}{ P_{M_{\rm th},\ell}(k_i)}  -1 \Big) P_{\ell}(k_i) {\rm Cov}^{-1} \Big[ \hat{P}_{\ell}(k_i), \hat{P}_{\ell^{'}}(k_j) \Big] \frac{\partial P_{\ell^{'}}(k_j)}{\partial p_{\beta}},\label{fisher_bias}
\end{align}
where $P^{\rm selection}$ and $P^{\rm w/o~selection}$ are the power spectra with and without the selection effect, and we have assumed that 
the spectra are given by $P_{\rm AM}$ and $P_{{M_{\rm th}}}$, respectively; 
$P_{M_{\rm th},\ell}$ is the $\ell$-th moment of the 
power spectrum, ${\rm Cov}^{-1}$ is the inverse of the covariance matrix, and 
$F^{\ell\ell'}$ is the Fisher matrix.
We assume the Gaussian covariance matrix and use the method in Refs.~\cite{2010PhRvD..81b3503G,2020PhRvD.101b3510K}
to analytically compute the auto- and cross-covariance matrices for the multipole moments of the power spectrum.
We use \texttt{Dark Emulator} to compute the $P_\ell$, the covariance matrix, 
and $F^{\ell \ell^{'}}$.
We include the information of the monopole and quadrupole moments, i.e. up to $\ell=2$, because 
the hexadecapole moment ($\ell=4$) is sub-dominant in the cosmological information content,
as shown in 
\cite{2020PhRvD.101b3510K}.
The quantity $\delta p_{\alpha}$ estimated from the above equation quantifies a bias in the parameter $p_{\alpha}$ due to
the selection effect, if the model prediction does not include it.
We here adopt $\mathbf{p} = \{ \Omega_m,~\sigma_8,~\alpha_{\parallel},~\alpha_{\perp},~M_{\rm th}, P_{\rm res}$\} as
a set of the model parameters, where $P_{\rm res}$ is the residual shot noise \cite{2003ApJ...598..720S} and $\alpha_{\parallel,~\perp}$ are the parameters to model the Alcock-Paczynski (AP) effect \citep{1979Natur.281..358A,2008PhRvD..77l3540P}.
By treating $\alpha_{\parallel,~\perp}$ as free parameters in the Fisher analysis, we can marginalize over the geometrical information, mainly from the BAO features, to estimate the selection effect in the $\sigma_8$ parameter,
which is constrained by the amplitude information of the power spectrum.
Throughout this analysis, the redshift is $0.5$, the same as the snapshot of the halo catalog we use in Section~\ref{sub:halo_catalog}.
The fiducial values of $M_{\rm th}$ we need for the calculation of Eq.~(\ref{fisher_bias}) are the thresholds we adopted to draw the red solid lines in Figs.~\ref{fig:results_LOWZ} and \ref{fig:results_CMASS}.
We also set other fiducial values of the parameters to the ones for the {\it Planck}~2018 cosmology, and assume survey volumes to be $1.98$ and $2.26~(h^{-1}\rm{Gpc})^3$ in the LOWZ and CMASS cases, respectively.

Fig.~\ref{fig:bias_in_estimation} shows the results for a possible bias in the estimation of $\sigma_8$, for the LOWZ- and CMASS-like samples.
The biases for other cosmological parameters, such as $\Omega_m$, $\alpha_{\parallel}$ and $\alpha_{\perp}$, are similar to the one for $\sigma_8$, so we omit them from the figure.
This figure shows that the selection effect does not cause an amount of bias in these parameter greater than the marginalized error, even when the maximum wavenumber of the analysis is $k_{\rm max}=0.3~h\rm Mpc^{-1}$.
Therefore, we conclude that the selection effect is unlikely to be significant if the selection effect is like $n_{\rm g}(z)$ 
of the LOWZ and CMASS samples.

\section{Conclusion}\label{sec:Conclusion}

In this paper, we have studied the impact of the selection effect of galaxies on the power spectrum. 
%the clustering analysis. 
To evaluate the galaxy selection effect, we mimicked the selection function by selecting halos above 
the redshift-dependent mass threshold in the N-body simulation, such that the resulting redshift distribution of the number density reproduces the target $n(z)$ for LOWZ- and CMASS-like galaxies. 
We demonstrated analytically that the selection effect inevitably introduces 
a bias in the redshift-space power spectrum (see Section~\ref{sec:linear_limit}).
We then quantified the impact of the selection effect by comparing the power spectra measured 
from the galaxy mock catalogs with those from the catalogs of single mass-threshold (and therefore 
single-tracer)
sample.
%we made two types of mock galaxy catalog, one of which includes the effect and the other does not.
%We measured the real- and redshift-space power spectrum from each type of mock catalog, and compared the two types of power spectra.
We found that 
%The results showed that 
the selection effect causes fractional changes of up to $1$ and $2\%$ 
in 
the monopole and quadrupole moments of the redshift-space power spectrum, up to $k=0.3~h{\rm Mpc}^{-1}$,
for the LOWZ- and CMASS-like galaxies. 
%$1\%$ up to the scale of $k\simeq 0.3h\rm Mpc^{-1}$, while can change the quadrupole $2\%$ up to the same scale, in the both cases of LOWZ and CMASS.
%We succeeded in explaining the behavior of the ratio $P_{M_{\rm th}}/P_{\rm AM}$ by introducing the local power spectrum  $P(\mathbf{k};\mathbf{x})$, and showed that these errors in monopole and quadrupole were unavoidable.
%We also studies the selection effect in the cases of power-law selection ($n(z)\propto z^{-\alpha}$), and found that the selection effect becomes stronger as the 
%resulting galaxy density shape becomes steeper.
Thus, our results imply that, for the {\em given} cosmological model, the halo mass dependencies of linear and nonlinear halo bias, nonlinear clustering, and the RSD effect are unlikely to cause significant bias in the redshift-space power spectrum, as long as the selection effect, quantified by
$n(z)$, does not have a significant redshift dependence. 
We estimated a possible bias in the $\sigma_8$ estimation
due to the neglect of the selection effect on the redshift-space power spectrum using the Fisher method, and showed that this effect does not cause a significant bias in $\sigma_8$ compared to the statistical error 
for a volume of $2~(h^{-1}{\rm Gpc})^3$.
%this effect hardly affects the cosmological parameter estimation.
This is good news for the current galaxy survey and will provide useful guidance for designing a wide-area spectroscopic galaxy survey.
However, this selection effect might need to be taken into account for ultimate galaxy surveys covering 
volumes larger than $\sim 50~(h^{-1}{\rm Gpc})^3$ such as the DESI and Euclid surveys. 

In this paper we worked on the halo catalogs, not galaxies. However the results of this paper are qualitatively applicable to galaxy clustering, because 
galaxies form in their halos and therefore galaxy clustering is given by a weighted sum of halo clustering over different halo masses, e.g. using the halo occupation distribution in halo model approach \citep{2019ApJ...884...29N,2022PhRvD.106h3519M,2022PhRvD.106h3520M,2020PhRvD.101b3510K,kobayashi2022fullshapecosmologyanalysissdssiii}. 
The exact magnitude  of the bias in the galaxy clustering amplitude due to the selection effect requires the use of a realistic mock catalog of galaxies for a sample of galaxies under consideration. We believe that the method developed in this paper would be useful for such a study, and can also be used when designing future galaxy surveys.

\acknowledgments
We would like to thank Yosuke~Kobayashi for useful
discussion and for allowing us to use the \texttt{Dark~Emulator} for the redshift-space power spectrum 
of halos
in this paper. 
We also thank 
Toshiki~Kurita for useful discussion. 
This work was supported in part by World Premier International Research Center Initiative (WPI Initiative), MEXT, Japan, 
and
JSPS KAKENHI Grant Numbers 
19H00677,
20H05850,
20H05855,
23KJ0747,
and 24H00215.
K.N. also sincerely acknowledges the financial support from the research assistant program
in AI \& Climate Data-Driven ELSI-RRI Study led by Prof.~Hiromi~Yokoyama.


\bibliographystyle{JHEP}
\bibliography{refs}
\end{document}